\documentclass[%
 preprint,
groupedaddress,
 amsmath,amssymb,
 aps,
pre,
]{revtex4-1}
\usepackage{enumerate}
\usepackage{graphicx}
\usepackage{dcolumn}
\usepackage{bm}
\usepackage{graphicx}
\usepackage{bbold} 
\usepackage{graphicx}
\usepackage{dcolumn}
\usepackage[left=3cm,top=2.5cm,right=3cm,bottom=2.5cm]{geometry}
\usepackage{array}
\usepackage{epsfig}
\usepackage{wrapfig}
\usepackage{color}
\usepackage{soul}
\usepackage{braket}
\usepackage{verbatim}
\usepackage{mathtools}
\usepackage{mathrsfs}
\usepackage{amsfonts}
\usepackage[english]{babel}
\usepackage{textcomp}
\usepackage{float}
\usepackage{comment}
\usepackage{url}
\usepackage{CJK}
\excludecomment{toexclude}
\usepackage{changepage} 

\begin{document}
\begin{center}
\textbf{The shape memory effect and minimal surfaces}\\
Mengdi Yin (\begin{CJK}{UTF8}{gbsn}尹梦迪\end{CJK}) and Dimitri D. Vvedensky\\
{\it The Blackett Laboratory, Imperial College London, London SW7 2AZ, UK}
\end{center}

\noindent{\textbf{Abstract:}} Martensitic transformations, viewed as continuous transformations between triply periodic minimal surfaces (TPMS), as originally proposed by Hyde and Andersson [Z. Kristallogr.~{\bf 174}, 225 (1986)], is extended to include paths between the initial and final phases. Bravais lattices correspond to particular TPMS whose lattice points are flat points, where the Gaussian curvature vanishes. Reversible transformations, which correspond to shape memory materials, occur only if lattice points remain at flat points on a TPMS throughout a continuous deformation. For the shape memory material NiTi, density-functional theory (DFT) yields irreversible and reversible paths with and without energy barriers, respectively. Although there are TPMS for face-centered ($\gamma$--Fe) and body-centered ($\alpha$--Fe) cubic lattices, $\gamma\to\alpha$ deformation paths are not reversible, in agreement with non-vanishing energy barriers obtained from DFT.\\


Martensitic transformations are diffusionless solid-to-solid transformations triggered by the rapid collective movement of atoms across distances smaller than a lattice spacing \cite{olson81,christian02,battacharya03}. This transformation is abrupt, showing a structural discontinuity at a particular temperature, and is often accompanied by changes in physical properties that can be exploited for applications. Materials exhibiting martensitic transformations include metals and alloys \cite{christian02}, ceramics \cite{kelly02}, and biological systems \cite{olson99}. 

The shape memory effect (SME) \cite{battacharya03,tadaki98}, where a material returns to its original shape when heated after a plastic deformation, accompanies some martensitic transformations. Shape memory materials have been known since the 1930s \cite{olander32}, but studies of this effect began in earnest only in the 1960s, with the discovery of NiTi-based materials \cite{buehler63,kauffman97}. Ni-Ti alloys are the most widely used shape memory and superelastic alloys, combining a pronounced SME with superelasticity, corrosion resistance, biocompatibility, and superior engineering properties.

Here, we study martensitic transformations from a geometric perspective with no restriction on point-group-subgroup relations between the two end phases to make our model appropriate for both reconstructive and non-reconstructive transformations. The martensitic transformation is interpreted as a continuous deformation between minimal surfaces, and includes information about the relevant symmetry and topological properties of the two end states. We build on the work of Hyde and Andersson \cite{hyde86} to develop a theory of martensitic transformations that encompasses not just the end-points but also the path between the initial and transformed phases. Transformation paths whose lattice points remain at flat points on surfaces in a space of triply periodic minimal surfaces (TPMS) are expected to be shape memory materials. Our approach provides a purely geometrical distinction between reversible and irreversible transformations.

The local shape near a point of a three-dimensional smooth surface is determined by the principal curvatures $k_1$ and $k_2$ measuring the maximum and minimum bending near that point. The geometry of a surface is characterized by the mean curvature $M=-{1\over2}(k_1+k_2)$ and the Gaussian curvature $K=k_1k_2$ \cite{gray97,oneill06}. A surface with $K=0$ everywhere is flat, e.g.~a plane.  A surface with $M=0$ everywhere is locally area-minimizing and is called minimal. Soap films, the catenoid (the surface of revolution of a catenary), and the plane are standard examples of minimal surfaces. Minimal surfaces with repeating units in three independent directions are called TPMS. Minimal surfaces can have points where $K=0$;~these are called flat points, and will play a central role in what follows. Flat points on the minimal surfaces considered here are isolated. 

TPMS can have space groups, point groups, just as ordinary crystals \cite{koch93}. The first TPMS were discovered by Schwarz \cite{schwarz90}, including the primitive ($P$), diamond ($D$), and hexagonal ($H$) surfaces. Schoen \cite{schoen70,schoen12} discovered the gyroid ($G$), which is an intermediate structure when deforming $P$ to $D$ surfaces along the Bonnet path and is locally isometric to both the Schwarz $P$ and $D$ surfaces \cite{schoen70,karcher89}. The $rPD$ family discovered by Schoen \cite{schoen70} and Karcher \cite{karcher89} is a set of TPMS connecting the $D$ and $P$ surface other than by the Bonnet transformation. Additional structural details of these surfaces are shown in the Supplementary Information. Since the point group of a lattice should be a subgroup, both proper and improper, of that of a surface \cite{kocian09}, we require the point group of the crystal to be a subgroup (proper and improper) of that of the TPMS at two end states to ensure the conformity between the surface and the crystal. While for intermediate states, because they are unstable, the group-subgroup relation, which may act as a main role in the geometric explanation to the twinning mechanism\cite{chen19}, between the surface and the lattice is not strictly required.  
\begin{figure}[t]
	\centering
	\includegraphics[width=0.8\linewidth]{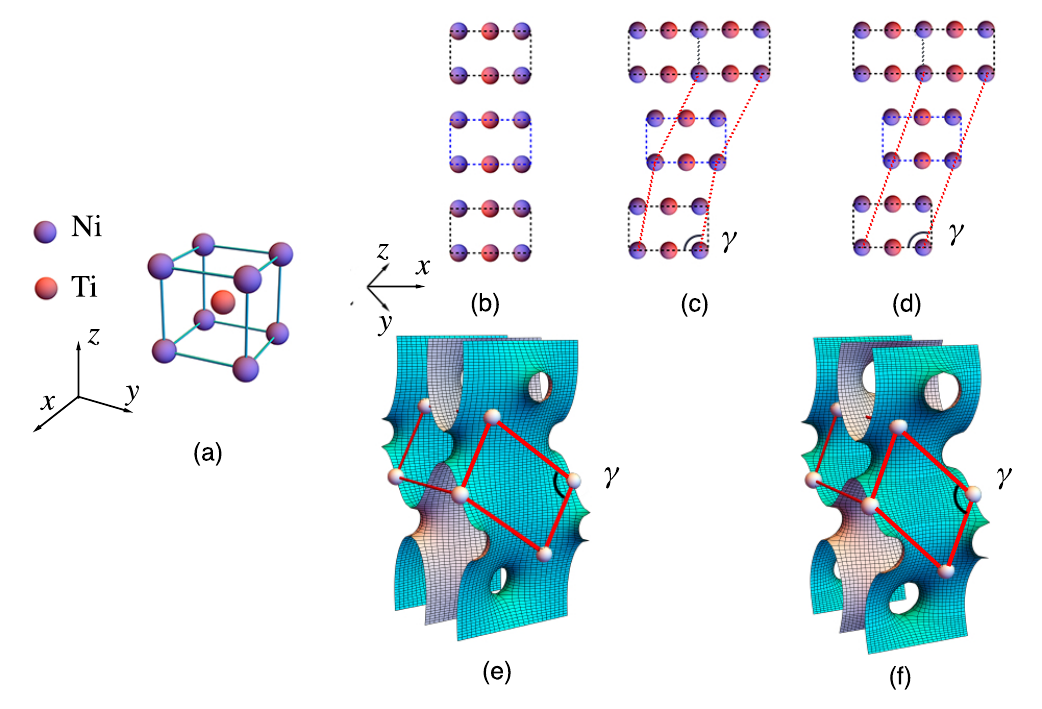}
	\caption{Deformation from B2 to B19$^\prime$ NiTi by bilayer $\langle100\rangle\{011\}$ shear. (a) Unit cell of NiTi. (b) View of undeformed B2 NiTi from the $[01\overline{1}]$ direction. B2 deformed by bilayer $\langle100\rangle\{011\}$ shear, with (c) $\gamma=98^\circ$ for B19$^\prime$ and (d) $\gamma=109^\circ$ for 109$^\circ$--B19$^\prime$. Red dashed lines outline the unit cell and blue lines label the bilayers. (e) and (f) Minimal surfaces corresponding to the B19$^\prime$ and 109$^\circ$--B19$^\prime$ lattices, drawn with the package Mesh \cite{weber18}, with white spheres labelling lattice points (not atoms).}
	\label{fig1}
\end{figure}

The austenite (B2) phase of NiTi has the space group $Pm\bar{3}m$. After a rapid quench, NiTi transforms to martensite. DFT calculations \cite{hatcher09,krcmar20,morris06} confirmed that there exist barrierless transformation paths of NiTi from the B2 phase to its martensite phases. We study NiTi structures along transformation paths by comparing the calculated and real ratios $a/c$ and $a/b$, where $a$, $b$, and $c$ are lattice parameters, and unit cell angles $\alpha$, $\beta$, $\gamma$ to ascertain if they can meet the requirements mentioned above. The calculated result is obtained by first constructing a ``unit cell'' from flat points representing lattice points rather than actual atoms of a multi-atom basis on the corresponding TPMS, then compute angles and ratios between cell edges. The actual lattice parameters can be obtained by a uniform scaling factor once the ratios are given. According to results listed in the Supplementary Information, all initial and final states calculated by DFT of shape-memory NiTi correspond to a TPMS.

DFT calculations \cite{hatcher09} also show that energy barriers vary with the paths of the B2$\to$B19$^\prime$ NiTi martensitic transformation, with the energy barrier going to zero when a path meets our geometrical criteria. Paths involving only the monolayer (resp., bilayer) $\langle 100\rangle\{011\}$ shear/shuffle have an energy barrier up to 37 meV/atom (resp., 22 meV/atom), while the path that admits the intermediate phase 109$^\circ$--B19$^\prime$, allowing both shear and structural relaxation, has no barrier, with the energy per atom decreasing throughout the transformation \cite{hatcher09}. Paths such as B2$\to$109$^\circ$--B19$^\prime$$\to$B19$^\prime$ correspond to a continuous deformation of the initial surface, where each surface along the path a TPMS (formally, in the oPb/oPa family \cite{fogden92,chen21}) and is, therefore, reversible. But for paths that allow only the bilayer $\langle100\rangle\{011\}$ shear, as shown in Fig.~\ref{fig1}(c), the co-existence of the B19$^\prime$ lattice and another lattice with $\gamma =119^\circ$ cannot form with flat points on the same TPMS. In other words, to describe the lattice in Fig.~\ref{fig1}(c),  there is a discontinuous transition surface consisting of both pieces of the minimal surfaces shown in Fig.~\ref{fig1}[(e,f)].  Thus, the surfaces are either non-minimal, or lattice points on a minimal surface depart from the flat points. Hence, according to our model, an energy barrier is obtained. 
\begin{figure}[t]
	\centering
	\includegraphics[width=0.7\linewidth]{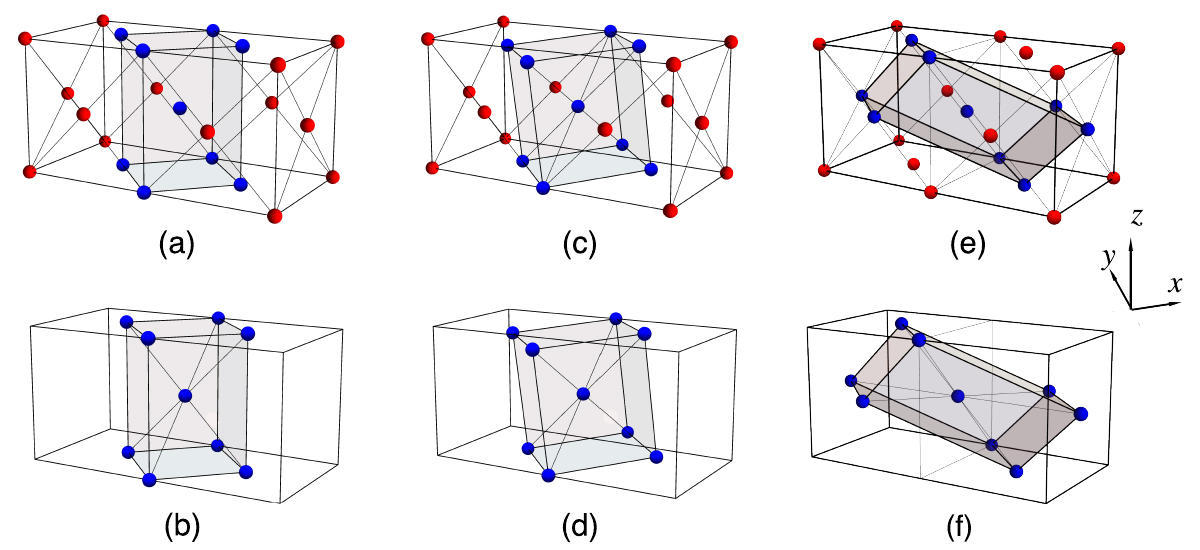}
	\caption{Atomic arrangements of (a) the Bain transformation path from the FCC to the BCC phase with an intermediate BCT unit cell (shown shaded), which is isolated in (b), (c) the NW path, with an intermediate body-centered triclinic unit cell (shown shaded), which is isolated in (d), (e) the KS transformation path, with an intermediate body-centered triclinic cell, as isolated in (f). Red and blue spheres indicate atoms in the parent and transformed phases, respectively.}
	\label{fig2}
\end{figure}

The best known and technologically most important martensite transformations are found in iron and steel \cite{pereloma12}. When cooled rapidly, high-temperature austenite $\gamma$--Fe (FCC)  transforms to martensite $\alpha$--Fe (BCC). Explanations for this transformation include the celebrated Bain tetragonal distortion \cite{bain24}, the Kurdjumow--Sachs (KS) \cite{sachs30} and Nishiyama--Wassermann (NW) \cite{nishiyama78,wassermann33} shear models, together with the Bogers--Burgers (BB) \cite{bogers64} and the Olson--Cohen (OC) \cite{olson72} hard-sphere models.~Where energies along transformation paths were calculated \cite{okatov09,zhang21}, a barrier of approximately 20~meV/atom was found for the $\gamma\to\alpha$ transformation of Fe.

\begin{figure}[t]
	\centering
	\includegraphics[width=0.9\linewidth]{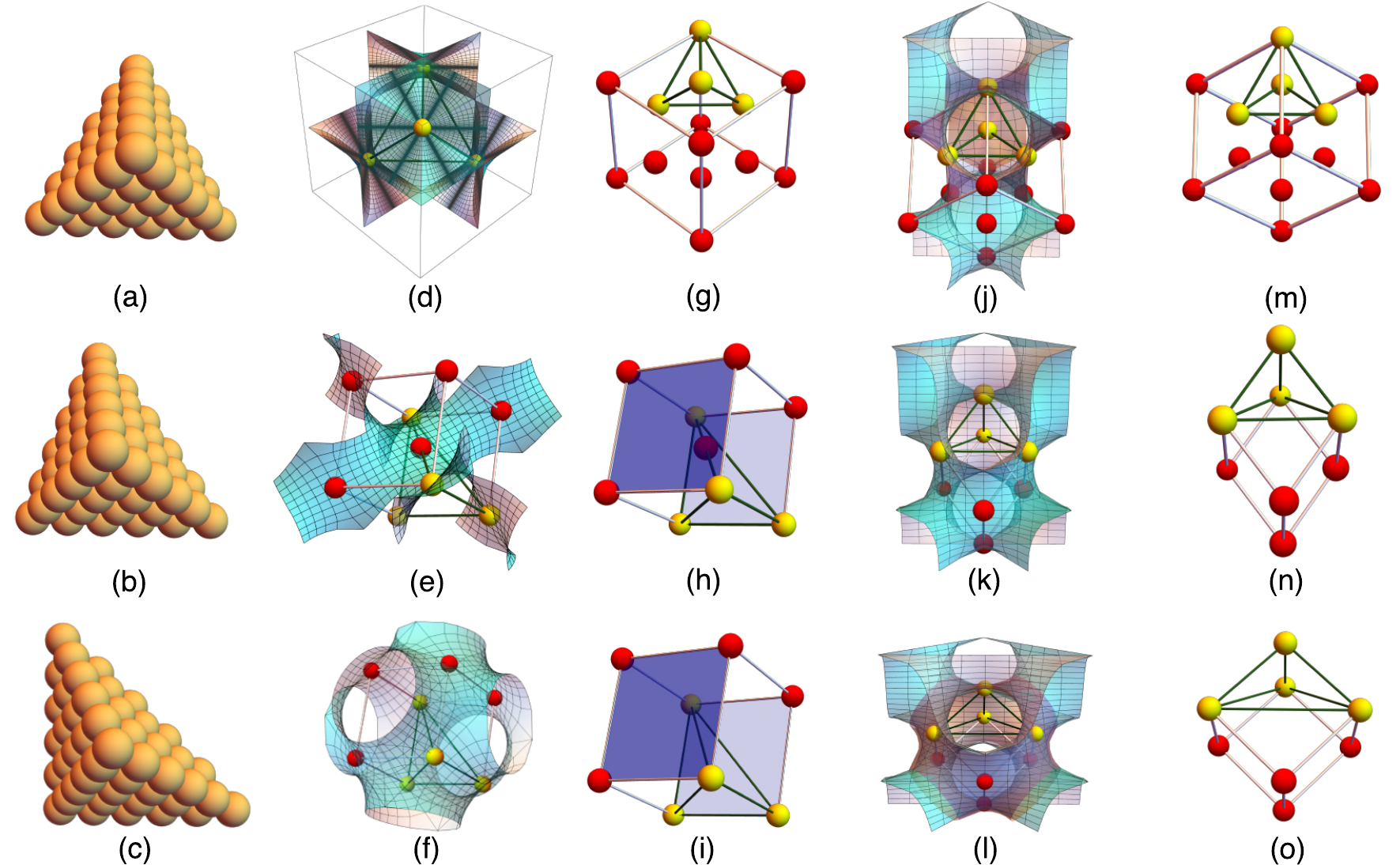}
	\caption{(a,b,c) Sketches of the BB (OC) hard-sphere model showing the stacking of $\{111\}$ planes. (a) Regular tetrahedron in the FCC lattice on the $D$ surface (d) and (j), and the irregular tetrahedron in (c) residing on either the gyroid (e) or the $P$ surface (f) and (l). (g,h,i) Ball-and-stick models of lattices formed on $D$, $G$ and $P$ surfaces, respectively. Spheres label lattice points that coincide with flat points. Yellow spheres are lattice points forming the vertices of tetrahedrons sketched by green lines. (j,k,l) depicts the $rPD$ path, with the area shading purple in (j) and (l) being (d) and (f), respectively. Formation of surfaces shown here are given in the Supplementary Information. (m,n,o) Ball-and-stick models of lattices formed along the $rPD$ path with the FCC lattice (m) on the $D$ surface and the BCC lattice (o) on the $P$ surface being two end states. (k) A representative of the $rPD$-family along the path, with the lattice shown in (n). Panels (d,e,f) and (j,k,l) were drawn with the package Mesh \cite{weber18}.}
	\label{fig3}
\end{figure}

The geometries for the Bain, NW and KS transformation paths connecting the FCC and BCC lattices are shown in Fig.~\ref{fig2}, while the BB (OC) path is shown in Fig.~\ref{fig3}(a,b,c). The Bain path [Fig.~\ref{fig2}(a,b)] is based on a deformation along $\langle001\rangle$, which yields a body-centered tetragonal (BCT) structure. The BCT unit cell is compressed by about 21\% along the $z$-direction and expanded by about 12\% along the $x$- and $y$-directions \cite{sandoval09}. Despite its simplicity and intuitive appeal, the orientation relationship between the parent (FCC) and daughter (BCC) phases has not been observed in experiments. The NW path [Fig.~\ref{fig2}(c,d)] begins with a shear that produces a monoclinic unit cell. Along this path, the unit cell changes from orthogonal to monoclinic and back to orthogonal. Computational studies \cite{zhang21,sandoval09} indicate this path is favored over the Bain path, though there is still a barrier for the transformation. The KS shear model [Fig.~\ref{fig2}(e,f)], similar to the NW model, starts with a shear that produces a monoclinic cell, but has different orientational relationships.  The BB (OC) hard-sphere model indicates the FCC-BCC lattice transformation is realized by a shear of type $\langle 112\rangle\{111\}$ \cite{olson72,bogers64}. 

The Bonnet and $rPD$-family of transformations are the only known continuous transformations that map the minimal surfaces for the BCC and FCC lattices to each other within the space of TPMS \cite{hyde86,fogden93,karcher89,schoen70}. The BCC lattice can form on Schoen's gyroid during the Bonnet transformation [Fig.~\ref{fig3}(d,e,f)]. This is an intermediate state, but due to the incompatibility of point group symmetry, this lattice cannot be stable on the $G$ surface. However, we can still expect the appearance of a BCC lattice during the Bonnet transformation. Whereas the $rPD$ path involves a vertical stretching of the surface [Fig.~\ref{fig3}(j,k,l)] and therefore involves the evolution of a tetrahedron with three faces of isosceles triangles and one face an equilateral triangle [Fig.~\ref{fig3}(m,n,o)]. Because no models mentioned above allow the appearance of a BCC lattice twice to conform to the Bonnet transformation, or the preservation of a tetrahedron required by the $rPD$ path described above then, according to our model, none of them are reversible. Currently, all $\gamma\to\alpha$ iron transformation paths \cite{sachs30,bogers64,olson72,cayron15} and relevant molecular dynamic calculations \cite{karewar18} are developed with the models discussed above, and none of them show geometric conformity to the Bonnet/$rPD$ path, yielding non-vanishing energy barriers \cite{meiser20}. Accordingly, we infer that the $\gamma\to\alpha$ iron martensitic phase transformation is geometrically irreversible;~hence, no SME can take place. 

\begin{table}
\caption{\label{table2}Calculated transformation paths of NiTi and Fe from the indicated initial and final phases. Energy barriers $\Delta E$ are in units of meV/atom.}
\begin{center}
\item[]\begin{tabular}{lcllc}
\toprule
\multicolumn{1}{c}{Material} & \multicolumn{1}{c}{Initial} & \multicolumn{1}{c}{Path} & \multicolumn{1}{c}{Final} & \multicolumn{1}{c}{$\Delta E$}\\
\hline
NiTi \cite{hatcher09}& B2 &$\langle 100\rangle\{011\}$ shear (basal shear) & B19$^\prime$&$\sim$37\\
& B2 & $\langle100\rangle\{011\}$ shear (bilayer shear) & B19$^\prime$&$\sim$22\\
& B2 &109$^\circ$--B19$^\prime$ & B19$^\prime$&0\\
NiTi \cite{hatcher09,vishnu10}& B2 & B19 & B19$^\prime$&0\\
NiTi \cite{niu16} & B2 &[111] elongation & R & 0\\
NiTi \cite{hatcher09,hatcher09a,morris06,krcmar20} & B2 & $\langle100\rangle\{011\}$ (bilayer shear and shuffle)  &B33 &0\\
NiTi \cite{kibey09} & B2 &$\langle\bar{1}10\rangle\{110\}$ shear & B19 &$\sim$13\\
Fe \cite{okatov09,zhang21} & $\gamma$-Fe & Bain (FCT) &$\alpha$-Fe &$\sim$20\\
& $\gamma$--Fe & Bain & $\alpha$--Fe & $\sim$45\\
Fe \cite{zhang21} &$\gamma$--Fe &Nishiyama--Wassermann (FCT) & $\alpha$--Fe & $\sim$37\\
Fe \cite{wang21} & $\gamma$--Fe & Bogers--Burgers & $\alpha$--Fe &$\sim$25\\
\hline
\end{tabular}
\end{center}
\end{table}

Table~\ref{table2} provides energy barriers calculated with DFT of materials undergoing martensitic transformations along several paths.~As explained above, NiTi deforms according to, for instance, the $\langle1\overline{1}0\rangle\{011\}$ shear and Fe deforms along the Bain or NW path do not satisfy the geometric requirements for the existence of a non-vanishing energy barrier, which pre-empts those paths exhibiting the SME. Paths with a vanishing energy barrier conform to the geometric requirements we propose for the SME. More details of the paths in Table~\ref{table2} are given in the Supplementary Information.

The details of the atomic-scale mechanisms responsible for martensitic transformations are important for a quantitative understanding of the transformation processes, including the calculation of energy barriers. Our approach, based on the transformations between minimal surfaces provides a conceptually simple description based on geometrical and topological principles. As with transformation paths such as proposed by Bain and NW, and according to the group-subgroup relation \cite{kaushik04}, no attempt is made to model the nucleation process, just to establish a relationship between the initial and final phases and, in our case, the intermediate states as well. 

In summary, we have studied the martensitic transformation and the SME from the perspective of the differential geometry of TPMS. We propose that martensitic transformations can be understood as continuous deformations between TPMS with the Bravais lattices of two end phases being at flat points, respectively. If a martensitic transformation continues within a space of TPMS, and all lattice points remain at flat points throughout the progress, we expect the occurrence of the SME. This method is relatively simple, general and less time-consuming in comparison to DFT, and thus can be a shortcut for the discovery of more novel shape memory alloys.

\bibliographystyle{unsrt}
\bibliography{paper.bib}

\end{document}